\documentclass[conference]{IEEEtran}
\IEEEoverridecommandlockouts
\usepackage{cite}
\usepackage{amsmath,amssymb,amsfonts}
\usepackage{algorithmic}
\usepackage{graphicx}
\usepackage{textcomp}
\usepackage{xcolor}
\usepackage{hyperref}
\def\BibTeX{{\rm B\kern-.05em{\sc i\kern-.025em b}\kern-.08em
    T\kern-.1667em\lower.7ex\hbox{E}\kern-.125emX}}
\begin{document}

\title{Minimize Web Applications vulnerabilities through the early Detection of CRLF Injection\
\thanks{Identify applicable funding agency here. If none, delete this.}
}

\author{\IEEEauthorblockN{MD Asibul Hasan}
\IEEEauthorblockA{\textit{Dept. of Computer Science and Engineering} \\
\textit{Southeast University}\\
\href{mailto:asifhasan12355@gmail.com}{asifhasan12355@gmail.com}}
\and

\IEEEauthorblockN{Md. Mijanur Rahman}
\IEEEauthorblockA{\textit{Dept. of Computer Science and Engineering} \\
\textit{Southeast University}\\
\href{mailto:mijanur.rahman@seu.edu.bd}{mijanur.rahman@seu.edu.bd}}
}

\maketitle

\begin{abstract}
Carriage return (CR) and line feed (LF), also known as CRLF injection is a type of vulnerability that allows a hacker to enter special characters into a web application, altering its operation or confusing the administrator. Log poisoning and HTTP response splitting are two prominent harmful uses of this technique. Additionally, CRLF injection can be used by an attacker to exploit other vulnerabilities, such as cross-site scripting (XSS). Email injection, also known as email header injection, is another way that can be used to modify the behavior of emails. The Open Web Application Security Project (OWASP) is an organization that studies vulnerabilities and ranks them based on their level of risk. According to OWASP, CRLF vulnerabilities are among the top 10 vulnerabilities and are a type of injection attack. However, CRLF vulnerabilities can also lead to the discovery of other high-risk vulnerabilities, and it fosters a better approach to mitigate CRLF vulnerabilities in the early stage and help secure applications against known vulnerabilities. Although there has been a significant amount of research on other types of injection attacks, such as Structure Query Language Injection (SQL Injection). There has been less research on CRLF vulnerabilities and how to detect them with automated testing. There is room for further research to be done on this subject matter in order to develop creative solutions to problems. It will also help to reduce false positive alerts by checking the header response of each request. Automated alerts from security systems can provide a quicker and more accurate understanding of potential vulnerabilities and can help to reduce false positive alerts. Despite the extensive research on various types of vulnerabilities in web applications, CRLF vulnerabilities have only recently been included in the research. Utilizing automated testing as a recurring task can assist companies in receiving consistent updates about their systems and enhance their security.
\end{abstract}

\section{Introduction}
Cyber security is primarily concerned with the protection of anything that is connected to the internet. This can be an application/software, network, device, etc. There are numerous types of vulnerabilities in applications, such as SQL injection, cross-site scripting (XSS), and local file inclusion (LFI), while network vulnerabilities may include denial of service (DoS) attacks, sniffing, and spoofing[3] [4]. To ensure cyber security, engineers must prioritize confidentiality, integrity, and availability, which are the three letters upon which the CIA triad stands. The goal of this research is to identify a specific application vulnerability. The cyber security industry is massive and consists primarily of two teams: one works for the company while the other works against the company, typically the intruder. It is crucial that everyone in the Software Development Life Cycle (SDLC) maintains the process, but due to a lack of understanding or high value, some organizations skip security testing. However, security testing checks whether the software is vulnerable to cyber attacks, test the impact of malicious activities, and determine the long-term success of the software.\\
The majority of researchers covered different types of vulnerability which is also dangerous for web applications as well as other software. But there is less research on CRLF injection vulnerability and has been one of the most dangerous vulnerabilities in recent years. Because this vulnerability was discovered newly that is why there is no details research about this vulnerability so there is a scope to improve CRLF detection. Researchers have only investigated the other types of vulnerabilities where attacks are mainly based on HTTP attacks [1]. This research only covers web application vulnerability which is a very big issue in developing a secure application[5]. CRLF vulnerability is not a common vulnerability like Cross site scripting or SQL injection [5][6] but this can lead to other vulnerabilities and expose the system to critical information.\\
If there is any vulnerability that discloses the company's inside information or exposes users’/customers’ information that will be a huge disaster for the company. The most widespread flaw in web applications is the injection, which includes SQL, HTML, CRLF, and other types of injection. Other flaws include XSS, broken access control, security misconfiguration, exposed sensitive data, inadequate attack protection, using components with known vulnerabilities, using unprotected APIs, local file inclusion, and broken authentication and session management [1][7]. To solve this problem, many researchers have tried several methods but less research has been carried out in the area of CRLF vulnerability detection or discovery, and this still remains one of the most critical vulnerabilities. This can expose system information and hackers can steal confidential data from applications. CRLF is not only a single vulnerability but this can also lead to some other vulnerabilities mainly injection type of vulnerabilities. \\
The majority of organization who is concerned about their security hire a security specialist to prevent security breaches, and a security engineer to check for vulnerabilities manually but this process takes so much time and reduce productivity. As cyber threats evolve, security engineers are increasingly tasked with threat modeling, penetration testing, and automation to proactively determine the level of vulnerability. This is focused on a critical injection of vulnerability CRLF. CRLF refers to Carriage Return and Line Feed. It’s an injection attack that could lead to XSS attack by doing that attacker can grab the user session and in some cases can accelerate privilege[8] [9]. XSS attacks are a type of injection, in which malicious scripts are injected into otherwise benign and trusted websites. XSS attacks occur when an attacker uses a web application to send malicious code, generally in the form of a browser side script, to a different end user through such an online application. The code for a web browser often takes the form of a JavaScript segment, but it can also be HTML, Flash, or any other type of code that the browser is capable of executing. [10][11]. XSS vulnerabilities normally allow an attacker to masquerade as a victim user, to carry out any actions that the user is able to perform and to access any of the user's data[6].\\
There are so many researches that have been done based on web application vulnerability as well as network vulnerability and threats. The majority of these are dangerous attacks that can take over the full system. But though CRLF vulnerability is a new kind of vulnerability that has not been explored by researchers, most especially the specific vulnerability. Some software works impeccably with CRLF vulnerability but they are paid applications. To solve this problem with CRLF vulnerability, this research will provide insight and a logical approach for those who want to work in this area of interest.
The first sectionof this study described its abstract. The second section  described the introduction of the study. The third section conveyed the literature review, followed by themethodology, and finally the conclusion of the research.
\section{Literature Review}

According to our studies, there has been a fair amount of research done on vulnerability management. Some of the research has been focused on injection-based attacks including SQL injection, HTML injection, and also code injection. A study on three major SQLi techniques was implemented on the educational and financial websites of Bangladesh and executes analysis web applications for figuring out the security condition [1]. But there was no mention of any CRLF vulnerability. Some case studies have been conducted on various types of vulnerabilities in some websites in Bangladesh. Additionally, some papers have explored automated and manual penetration testing in a range of domains. An example of the online application called Tunestore is used in a case study to carry out security testing. It provides an example of tool- and manually-assisted web application security testing. Testing on Tunestore is done using Paros, WebScarab, JBroFuzz, Fortify, and Acunetix. [5]. \\
This paper aims to eliminate CRLF vulnerability on web applications and helps the security tester to detect the vulnerability before releasing the product. Solving this vulnerability will also secure the application against XSS attacks because CRLF can also lead to XSS. These are major vulnerabilities according to OWASP.

\section{Methodology}
CRLF vulnerability in web applications is a major security concern that can have serious consequences. This vulnerability allows attackers to insert malicious code into a web page or application, which can then be executed by the web browser or program. This can result in the exposure of sensitive information, the execution of arbitrary code, or the launch of a denial of service attack.\\
CRLF vulnerabilities are often exploited through CRLF injection attacks, in which malicious code is injected into a web page or application. To prevent CRLF injection attacks, it is essential to properly validate and sanitize all input. Any user input that will be used in a Structured Query Language (SQL)  should be properly encoded and checked for incorrect characters. It is also critical to keep all web servers and applications up to date with the latest security fixes. \\
\begin{center}
    \includegraphics[width=\linewidth]{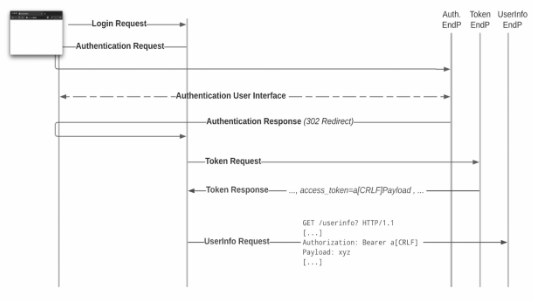}
    Figure 01: How CRLF attacks occurred
\end{center}

As a result of this CRLF attack, more harmful attacks including XSS, page injection, web cache poisoning, and many others are launched. Log poisoning and HTTP response splitting are the popular use of these attacks. By adding a line end and an extra line, the attacker adds false log file entries. This could be done to deceive system administrators or cover up other attacks [10]. LF, CR, \#, and ! are common ASCII characters used in creating server-side attacks. By including them in the feature set, these assaults can be detected.[2] \\
One way to exploit a CRLF vulnerability is to inject a CRLF character into a web application in order to exploit a buffer overflow vulnerability. Another way to exploit a CRLF vulnerability is to inject a CRLF character into a web application in order to exploit an XSS vulnerability.
Proposed framework to find CRLF vulnerability: 

\begin{center}
    \includegraphics[width=\linewidth]{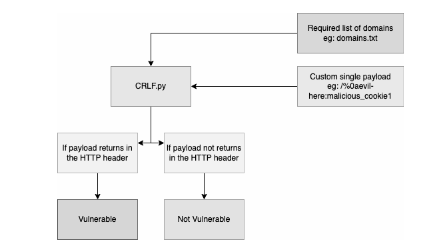}
    Figure 02: Proposed Framework
\end{center} 
In this figure, the user will give a list of website links or a single link when running CRLF. After that, the application will check for header responses if there is CR or LF signs based on that application and will make sure whether it is vulnerable or not vulnerable.
This framework will give fewer false positive alerts than other applications. 
An extract of the complete HTTP GET request is shown below:[1]
\begin{center}
    \includegraphics[width=\linewidth]{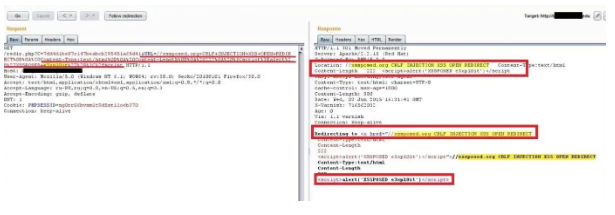}
    Figure 03: CRLF in the header
\end{center} 
In this figure, it is a header request where CRLF means the CR and LF tag can be found. \\
CR and LF are special characters (ASCII 13 and 10 respectively, also referred to as $\backslash$r and $\backslash$n) that are used to signify the End of Line (EOL). They’re used to note the termination of a line, however, dealt with differently in today’s popular Operating Systems.
\section{Result and Discussions}
Our study concentrated on determining the presence and consequences of CRLF vulnerabilities in the wild as well as investigating potential remedies and the most effective methods for avoiding and overcoming these problems. Our research shows that CRLF vulnerabilities affect a large number of websites and online apps and that they are rather widespread in web applications. These flaws might have detrimental effects, such as allowing hackers to insert malicious code into a website or application, which could result in data breaches, identity theft, and other security breaches. We advise using a number of quality standards, such as input validation, sanitization, and encoding of user input, as well as routine testing and monitoring of web applications to discover and resolve any vulnerabilities, in order to mitigate these issues. Overall, our research emphasizes how critical it is to handle online security in a proactive manner, including routinely identifying and patching possible vulnerabilities like CRLF issues. By doing this, businesses can defend themselves against security flaws, guarantee the security of their users, and safeguard their websites and apps.
\begin{center}
    \includegraphics[width=\linewidth]{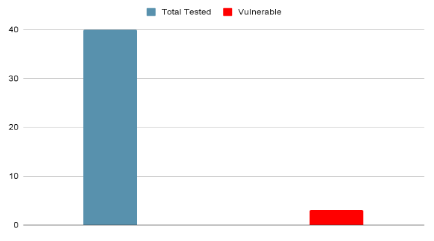}
    Figure 04: Vulnerability chart
\end{center} 
The findings of this investigation showed that a total of 40 websites were examined using the suggested framework for locating CRLF vulnerabilities, which are shown in Figure 4. Three out of the forty websites were determined to be vulnerable to CRLF injection attacks, according to the data graph. 
For some legal issues, it’s not possible to disclose the target website’s name or website address.
\begin{center}
    \includegraphics[width=\linewidth]{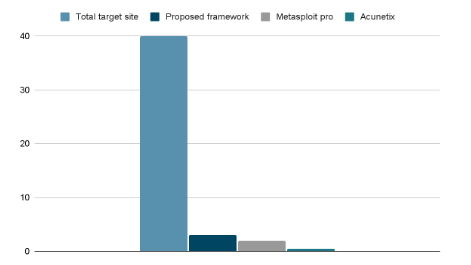}
    Figure 05: Comparison chart
\end{center} 
The findings of the research showed that 40 websites were examined for CRLF vulnerabilities using three distinct frameworks which are depicted in Figure 5. According to the statistics in the figure, the suggested framework was able to find more susceptible websites than Acunetix and Metasploit Pro combined.
\section{Conclusion}
This study's objective was to better our knowledge about CRLF vulnerabilities in web applications. This study examined the characteristics and potential repercussions of CRLF vulnerabilities as well as techniques for spotting and reducing these dangers. The suggested framework was more effective than the already available tools and had fewer false positives. Our research has shown the significance of taking CRLF vulnerabilities into account during the software development lifecycle and highlighted how they could affect the security of online applications. By offering practical knowledge that may assist people and organizations in defending against the continuously changing threats in the digital world, this study has also contributed to the larger area of cyber security.

\end{document}